\documentclass[preprint,aps,prb,amsmath]{revtex4}
\usepackage{latexsym}
\usepackage{amssymb}
\usepackage{graphicx}
\usepackage{bm}% bold math

\newcommand{\pdag}{{\phantom{\dagger}}}
\newcommand{\bq}{\begin{equation}}
\newcommand{\eq}{\end{equation}}
\newcommand{\bn}{\begin{eqnarray}}
\newcommand{\en}{\end{eqnarray}}

\begin{document}

\title{Super-Poissonian shot noise in the resonant tunneling due to coupling with a %%@
localized level}

\author{Ivana Djuric}
\affiliation{Department of Physics and Engineering Physics, Stevens Institute of %%@
Technology, Hoboken, New Jersey 07030}
\author{Bing Dong}
\affiliation{Department of Physics and Engineering Physics, Stevens Institute of %%@
Technology, Hoboken, New Jersey 07030 \\
Department of Physics, Shanghai Jiaotong University, 1954 Huashan
Road, Shanghai 200030, China}
\author{H. L. Cui}
\affiliation{Department of Physics and Engineering Physics, Stevens Institute of %%@
Technology, Hoboken, New Jersey 07030 \\
School of Optoelectronics Information Science and Technology, Yantai University, Yantai, %%@
Shandong, China}

\begin{abstract}

We report our studies of the shot noise spectrum in tunneling through an interacting %%@
quantum dot when an additional single-level quantum dot without tunnel coupling to leads %%@
is coherently side-connected to it. We show that the zero-frequency shot noise could %%@
reach a super-Poissonian value for appropriate ratios between dot-dot hoppings and %%@
dot-lead couplings, but the current is independent on the hopping.
Moreover,
the frequency spectrum of shot noise shows an obvious peak at the Rabi frequency, which %%@
is controllable by tuning the dot-lead couplings.

\end{abstract}


\maketitle

A measurement of the stationary I-V characteristic does not always
provides enough information to describe a charge transport
mechanism through a mesoscopic system. Shot noise,\cite{Blanter}
i.e. fluctuation of the current in time due to the discrete nature
of electrons, characterizes the degree of correlation between
charge transport events and it can be used as an additional
diagnostic tool to distinguish various transport mechanisms
possibly resulting in the same mean current.

The resonant tunneling (RT) through a single localized state
(quantum dot) has been a subject of intensive investigations. It
has been found that the current and the Fano factor, which
quantifies correlations with respect to the uncorrelated
Poissonian noise, are given by the relations:\cite{Chen,Nazarov}
\bn I=e\frac{\Gamma_{L}\Gamma_{R}}{\Gamma_{L}+\Gamma_{R}},
\label{I0} \en and \bn F=S_{I}/2eI=1-
\frac{2\Gamma_{L}\Gamma_{R}}{(\Gamma_{L}+\Gamma_{R})^2}.
\label{F0} \en Here, $S_{I}=2\int_{-\infty}^{\infty}dte^{i\omega
t}\left[\langle I(t)I(0)\rangle-\langle I\rangle^{2}\right]$ is a
current noise power spectrum and $\Gamma_{L(R)}$ are the tunneling
rates between resonant level and left (right) reservoir. The Fano
factor has a values from $0.5$ for symmetric coupling
($\Gamma_{L}=\Gamma_{R})$ to $1$ for significantly different
rates. Suppression of shot noise ($F<1$ for asymmetric coupling)
is caused by the Pauli exclusion principle which forbids the
tunneling of an electron from reservoir into the resonant level as
long as the resonant level is occupied. This results in negative
correlations between two consecutive electron transfers and,
therefore, suppression of shot noise. On the other hand, Coulomb
interactions between particles may decrease or increase noise
correlations, depending on the physical regimes.

In this work we present a theoretical study of time-dependent
fluctuations of the RT current through a two tunnel-coupled
quantum dots (QD) in the Coulomb blockade regime when only one of
the QD is connected with the leads. This theoretical model, as
depicted in Fig.~1, can describe a RT through the quantum well in
presence of an impurity inside the well or a RT through the QD
when a conducting level is tunnel-coupled with the other localized
level of the QD. The shot noise is studied within a bias-voltage
and temperature dependent quantum rate equation\cite{Dong}
approach from our earlier paper.\cite{I}

Both systems can be described by the usual model Hamiltonian for
two coupled QDs (CQD): \bn H &=& \sum_{\eta, k}\epsilon _{\eta k}
c_{\eta k }^{\dagger }c_{\eta k}^{\pdag} + \epsilon_{A}c_{A}^{\dagger }c_{A}^{\pdag}+ %%@
\epsilon_{B} c_{B}^{\dagger }c_{B}^{\pdag} + %%@
Uc_{A}^{\dagger}c_{A}^{\pdag}c_{B}^{\dagger}c_{B}^{\pdag} \cr
&& + t\ (c_{A}^{\dagger} c_{B}^{\pdag} + c_{B}^{\dagger} c_{A}^{\pdag})+ \sum_{\eta, k} %%@
(V_{\eta} c_{\eta k }^{\dagger }c_{A}^{\pdag} +{\rm {H.c.}}),
\label{hamiltonian} \en
where $c_{\eta k }^{\dagger }$$(c_{\eta k})$ are the creation (annihilation) operators %%@
for electron with momentum $k$ and energy $\epsilon _{\eta k}$ in
the lead $\eta$ ($\eta=L, R$), $c_{A(B)}^{\dagger}$ ($c_{A(B)}$)
are creation (annihilation) operators for an electron in QD A(B),
respectively. $\epsilon_{i}$ ($i=$ A, B) is the bare energy level
of
electrons in the $i$th QD. Here we assume that only one bare energy level in each dot is %%@
involved in transport. The intradot electron-electron Coulomb interactions are assumed %%@
to be infinite but the interdot interaction $U$ is finite. Namely, the state of two %%@
electrons occupied in the same QD is forbidden but two electrons dwelling in different %%@
QDs is permitted. The $V_{\eta}$ describes the coupling between the QD A and lead %%@
$\eta$.

Under the assumption of weak coupling between the QD and the
leads, and applying the wide band limit in the two leads,
electronic tunneling through this system in sequential regime can
be described by the bias-voltage and temperature dependent quantum
rate equations for the dynamical evolution of the density matrix
elements, $\rho_{ij}(t)$.\cite{Dong} The statistical expectations
of the diagonal elements of the density matrix, $\rho_{ii}$
($i={0, 1, 2, d}$), give the occupation probabilities of the
states of the dots, namely: $\rho_{00}$ is the probability of
finding both dots unoccupied, $\rho_{11}$, $\rho_{22}$,
$\rho_{dd}$ are the probabilities of finding dot A, dot B, and
both of two dots occupied, respectively. The off-diagonal density
matrix elements $\rho_{12}=\rho_{21}^{\ast}$ describe the coherent
superposition of the two levels in different QDs. The resulting
quantum rate equations are:
\begin{subequations}
\label{rateq1} \bn
\dot{\rho}_{00}&=& (\Gamma_{L}^{-}+ \Gamma_{R }^{-}) \rho_{11} - (\Gamma_{L}^{+} + %%@
\Gamma_{R}^{+}) \rho_{00}, \label{rc01} \\
\dot{\rho}_{11}&=& (\Gamma_{L}^{+}+\Gamma_{R}^{+}) \rho_{00} - %%@
(\Gamma_{L}^{-}+\Gamma_{R}^{-}) \rho_{11}\cr &&+it( \rho_{12}-\rho_{21}), \label{rc11} %%@
\\
\dot{\rho}_{22}&=& (\widetilde{\Gamma}_{L}^{-}+\widetilde{\Gamma}_{R}^{-}) %%@
\rho_{dd}-(\widetilde{\Gamma}_{L}^{+}+\widetilde{\Gamma}_{R}^{+}) \rho_{22} \cr %%@
&&-it(\rho_{12}-\rho_{21}), \label{rc21} \\
\dot{\rho}_{12}&=& i(\epsilon_{1}-\epsilon_{2})+it (\rho_{11}-
\rho_{22}) \cr
&& - \frac{1}{2} [\Gamma_{L}^{-} + \Gamma_{R}^{-} + %%@
\widetilde{\Gamma}_{R}^{+}+\widetilde{\Gamma}_{L}^{+}] \rho_{12},
\label{rc31} \\
\dot{\rho}_{dd} &=& (\widetilde{\Gamma}_{L}^{+}+\widetilde{\Gamma}_{R}^{+}) \rho_{22} - %%@
(\widetilde{\Gamma}_{L}^{-} + \widetilde{\Gamma}_{R}^{-})
\rho_{dd}, \label{rc41} \en
\end{subequations}
together with the normalization relation
$\rho_{00}+\rho_{11}+\rho_{22}+\rho_{dd}=1$. The
temperature-dependent tunneling rates are defined as
$\Gamma_{\eta}^{\pm}=\Gamma_{\eta}f_{\eta}^{\pm}(\epsilon_{1})$
and
$\widetilde{\Gamma}_{\eta}^{\pm}=\Gamma_{\eta}f_{\eta}^{\pm}(\epsilon_{1}+U)$,
where $\Gamma_{\eta}$ are the tunneling constants,
$f_{\eta}^{\pm}(\omega)=\left[1+e^{(\omega-\mu_{\eta})/k_{B}T}\right]^{-1}$
is the Fermi distribution function of the $\eta$ lead and
$f_{\eta}^{-}(\omega)=1-f_{\eta}^{+}(\omega)$. Here,
$\Gamma_{\eta}^{+}$ ($\Gamma_{\eta}^{-}$) describes the tunneling
rate of electrons into (out of) the QD A from (into) the $\eta$
lead without the occupation of QD B, while
$\widetilde{\Gamma}_{\eta}^{+}$ ($\widetilde{\Gamma}_{\eta}^{-}$)
stands for the corresponding rates due to the Coulomb repulsion,
when the QD B is already occupied by an electron. The current
flowing through the system is given by: \bn
I_{L} = \Gamma_{L}^{+} \rho_{00} + \widetilde{\Gamma}_{L}^{+}\rho_{22} - %%@
\Gamma_{L}^{-}\rho_{11} + \widetilde{\Gamma}_{L}^{-}\rho_{dd}.
\label{I} \en In order to calculate the noise power spectrum we
employed the procedure from Ref.~[\onlinecite{I}].
In the basis $(\rho_{11}, \rho_{22}, \rho_{00}, \rho_{12}, \rho_{21}, \rho_{dd})$, the %%@
current operator, $\hat{\Gamma}_{L}$, has a matrix form with
non-zero elements: $(\hat{\Gamma}_{L})_{13}=\Gamma_{L}^{+}$,
$(\hat{\Gamma}_{L})_{31}=-\Gamma_{L}^{-}$,
$(\hat{\Gamma}_{L})_{25}=-\tilde{\Gamma}_{L}^{-}$ and
$(\hat{\Gamma}_{L})_{52}=\tilde{\Gamma}_{L}^{+}$.

In the following calculations we set the hopping $t=1$ as the unit
of energy, and
$\epsilon_{A}=\epsilon_{B}=1$, the Coulomb interaction $U=5$, and the %%@
thermal  energy $kT=0.1$. If $t$ is 1 $\mu$eV, then $\Gamma\sim
\mu$eV, and $kT=0.1$ meV, which correspond to a typical
experimental situation\cite{Exp}. The Coulomb interaction between
the two dot, for this set of parameters, would be $U=5$ $\mu$eV.
The zero of energy is chosen to be the Fermi level of the leads in
the equilibrium condition ($\mu_{L}=\mu_{R}=0$). The bias voltage,
$V$, between the source and the drain is considered to be applied
symmetrically, $\mu_{L}=-\mu_{R}=eV/2$.

In Fig.~2, we plot the calculated current and Fano factor as functions of bias-voltage %%@
for different dot-lead couplings $\Gamma=4$, $2$, $1$, and $0.5$.
At small bias, $eV\ll kT$, the noise is dominated by thermal noise
which leads to the divergence $2kT/eV$ of the Fano factor
occurring at $V=0$. When Fermi level of the source, $\mu_{L}$, is
well below $\epsilon_{A}$ the transport is exponentially
suppressed because there are very few tunneling events. In this
region, electron transport is limited only by thermally activated
tunneling, thus, tunneling events are uncorrelated and Fano factor
is Poissonian. The transport through the system becomes
energetically allowed when the Fermi level of the source,
$\mu_{L}$, crosses the discrete level $\epsilon_{A}$ (for
$eV/2\approx\epsilon_{A}$). In this region current starts to
increase from zero to a constant value $I_{1}$ (the first current
plateau): \bn
I_{1}=e\frac{\Gamma_{L}\Gamma_{R}}{\Gamma_{R}+2\Gamma_{L}},
\label{I1} \en which is independent on the hopping between two
QDs. In contrast to the
current, Fano factor $F_1$ (Fano factor corresponding to the first current plateau) is %%@
dependent on the hopping $t$: \bn
F_{1}=1-\frac{4\Gamma_{L}\Gamma_{R}}{(\Gamma_{R}+2\Gamma_{L})^2}+\frac{\Gamma_{L}^{2} %%@
\Gamma_{R}^{2}}{2t^{2}\left(\Gamma_{R}+2\Gamma_{L}\right)^{2}}.
\label{F1} \en Thus, the valuable information can be obtain
through the shot noise measurement. The second term in
Eq.~(\ref{F1}) describes the suppression of the Fano factor below
unity, whereas the third term gives a positive contribution. The
electron flow through the system is possible by the two different
paths: the direct path ($L \rightarrow$
A $\rightarrow R$) and the indirect one ($L \rightarrow$ A $\rightarrow$ B $\rightarrow$  %%@
A $\rightarrow R$). For weak coupling between QD A and leads, i.e.
$\Gamma_{L,R}\ll t$, electron performs fast oscillation between
two levels with the Rabi frequency
$\omega\sim 2t$. In such case, electron flowing through the indirect path has the same %%@
contribution to tunneling as electron through the direct path and the system becomes two %%@
interacting tunneling levels model. Clearly, the Fano factor is reduced to %%@
$F_1=1-\frac{4\Gamma_{L} \Gamma_{R}}{(\Gamma_{R} + 2\Gamma_{L})^2}$. On the contrary, in %%@
the case of strong coupling ($\Gamma_{L,R}\gg t$), the difference between two paths is %%@
distinct: electron flowing through the direct path has a fast characteristic time $\sim %%@
\Gamma_{L,R}^{-1}$, while electron tunneling through the indirect
path has a slow characteristic time $\sim (2t)^{-1}$. In the limit
where the Coulomb interactions prevent a double occupancy of the
central region, there is competition between these two path.
Consequently, the slow flowing of electrons through the indirect
path modulate the transport through the direct path, which leads
to a bunching of tunneling events and results in super-Poissonian
noise, as shown in Fig.~1(b). The super-Poissonian noise as a
consequence of the partial blockade of an electronic channel by
another one has also been found and described in
Ref.~[\onlinecite{Cottet}].

The second step in current-voltage characteristic occur when the
Fermi level $\mu_{L}$ crosses $\epsilon_{d}+U$, i.e. when the system becomes doubly %%@
occupied. In this region, the current $I_2$ and the Fano factor
$F_2$ reduce to the results
for one-level system coupled to the leads, Eqs.~(\ref{I0}) and (\ref{F0}), respectively. %%@
This result can be easily understood as follows: An electron
becomes "stuck" in dot B and makes this dot be effectively
disconnected from the system.

This theoretical model can be used to explain the experimental
results obtained in Refs.~[\onlinecite{Nauen}] and
[\onlinecite{Safonov}], where the investigation of the noise
properties of resonant transport through an impurity situated
within the quantum well of a tunneling structure has been
performed. The measured current-voltage characteristic show a
pronounced current step with a Fano factor
$F\approx 0.6$ and an additional second weak structure with $F>1$. Our studies claim %%@
that the two-step $I$-$V$ curve and super-Poissonian noise can be
caused by a localized level coupled with the conducting one.

More information about this system could be obtained through the
frequency-dependent shot noise studies. The frequency-dependent
Fano factor in the Coulomb blockade regime
($eV/2<\epsilon_{d}+U$), for different couplings $\Gamma_{R}$ and
$\Gamma_{L}$, is plotted in
Fig.~3. It is observed that the Fano factor displays an obvious peak at the %%@
characteristic frequency $\omega=2t$, because the Rabi oscillation has the most %%@
pronounced effect on enhancing the deviation of instant current from its average value %%@
at this value of frequency. For small coupling between QD A and
the right reservoir, $\Gamma_{R}<2t$, there is large probability
for an electron inside the CQD to tunnel back and forward many
times between two levels before it exits into the right lead. This
causes a remarkably sharp peak in the Fano factor at the Rabi
frequency $\omega=2t$. With increasing $\Gamma_{R}$, the
probability for an electron to tunnel out increases, which plays a
role to destroy oscillating between two levels and induces
broadening and decreasing the amplitude of the peak. On the other
hand, increasing $\Gamma_{L}$ increases the probability to have an
electron inside the system, leading to significant enhancement of
shot noise.

In conclusion, we have analyzed the shot noise properties of
resonant tunneling through coupled QDs when only one of the dots
is coupled to the reservoirs. The additional current step and the
super-Poissonian shot noise in this region have been found. We
have further suggested that experimental results in
Ref.~[\onlinecite{Nauen}] can be explained through this model. We
have also suggested that frequency dependent noise measurement
performed on this system could provide more information about
hopping strength between levels.

This work was supported by the DURINT Program administered by the US Army Research %%@
Office.

\newpage

\centerline{\Large Figure Caption}

\vspace{2cm}

\noindent FIG.1: Schematic picture of the system.
Quantum dot A is connected to the leads L and R via tunneling junctions with rates %%@
$\Gamma_{L}$ and $\Gamma_{R}$, respectively. It is also coherently coupled to a %%@
localized level in quantum dot B with a hopping rate $t$.

\vspace{1cm} \noindent FIG.2: Current (a) and Fano factor (b) vs.
the bias-voltage
in the CQDs calculated for different dot-lead couplings %%@
$\Gamma_{L}=\Gamma_{R}=\Gamma=4$, $2$, $1$, and $0.5$. Other
parameters are: $\epsilon_{d}=1$, $U=5$, $kT=0.1$.

\vspace{1cm} \noindent FIG.3: Fano factor vs. frequency in the
Coulomb blockade regime calculated for different couplings
$\Gamma_{R}$ (a) and $\Gamma_{L}$ (b). Other parameters are:
$\Gamma_{L}=1$ (a), $\Gamma_{R}=1$ (b), $\epsilon_{d}=1$, $U=5$,
$V=6$, $kT=0.1$.

\newpage

\begin{figure}[htb]
\includegraphics[height=2in,width=3in]{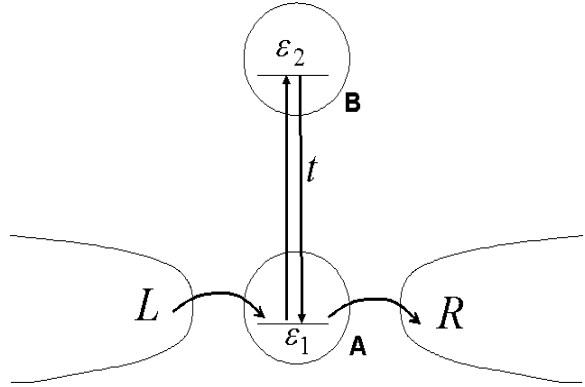}
\caption{Schematic picture of the system.
Quantum dot A is connected to the leads L and R via tunneling junctions with rates %%@
$\Gamma_{L}$ and $\Gamma_{R}$, respectively. It is also coherently coupled to a %%@
localized level in quantum dot B with a hopping rate
$t$.}\label{FIG.1}
\end{figure}

\newpage

\begin{figure}[htb]
\includegraphics[height=3in,width=3in]{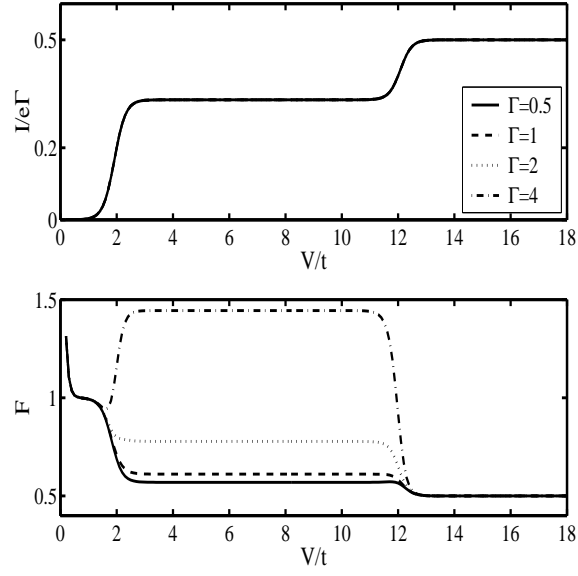}
\caption{Current (a) and Fano factor (b) vs. the bias-voltage
in the CQDs calculated for different dot-lead couplings %%@
$\Gamma_{L}=\Gamma_{R}=\Gamma=4$, $2$, $1$, and $0.5$. Other
parameters are: $\epsilon_{d}=1$, $U=5$, $kT=0.1$.}
\end{figure}

\newpage

\begin{figure}[htb]
\includegraphics[height=2in,width=3.4in]{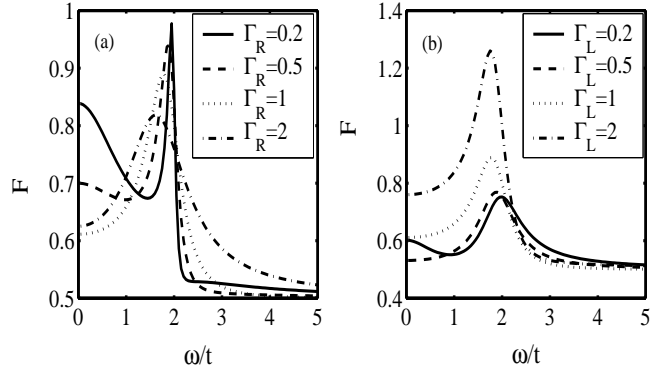}
 \caption{Fano factor vs. frequency
in the Coulomb blockade regime calculated for different couplings
$\Gamma_{R}$ (a) and $\Gamma_{L}$ (b). Other parameters are:
$\Gamma_{L}=1$ (a), $\Gamma_{R}=1$ (b), $\epsilon_{d}=1$, $U=5$,
$V=6$, $kT=0.1$.}
\end{figure}

\end{document}